\documentclass[manuscript,nonacm]{acmart} 
\AtBeginDocument{%
  \providecommand\BibTeX{{%
    \normalfont B\kern-0.5em{\scshape i\kern-0.25em b}\kern-0.8em\TeX}}}

\copyrightyear{2024}
\acmYear{2024}

\newcommand{\WOzTool}[0]{\textit{PilotAR}}

\newif\ifcomment
\commentfalse

\newif\ifrevision
\revisionfalse

\ifcomment
\ifrevision
\newcommand{\hide}[1]{}
\newcommand{\note}[1]{}
\newcommand{\added}[1]{\textcolor[rgb]{1,0,0}{#1}}
\newcommand{\modif}[1]{\textcolor[rgb]{0.9,0.3,0.9}{#1}}
\newcommand{\cut}[1]{}

\newcommand{\todo}[1]{}

\newcommand{\deleted}[1]{\textcolor[rgb]{0.6,0.6,0.6}{{#1}}}

\newcommand{\temporary}[1]{}

\newcommand{\nuwan}[1]{}
\newcommand{\runze}[1]{}
\newcommand{\zsd}[1]{}
\newcommand{\ashwin}[1]{}

\else

\newcommand{\hide}[1]{}
\newcommand{\note}[1]{\textcolor{blue}{<< #1 >>}}
\newcommand{\added}[1]{\textcolor[rgb]{0.1, 0.56, 1}{#1}}
\newcommand{\modif}[1]{\textcolor[rgb]{0.5,0.5,0.9}{#1}}
\newcommand{\cut}[1]{\textcolor[rgb]{0.5,0.5,0.5}{CUT: #1}}

\newcommand{\todo}[1]{\textcolor{red}{TODO: #1}}

\newcommand{\deleted}[1]{\textcolor[rgb]{0.7,0.7,0.7}{{#1}}}

\newcommand{\temporary}[1]{\textcolor[rgb]{0.9,0.9,0.9}{{#1}}}

\newcommand{\nuwan}[1]{\textcolor{blue}{NUWAN: #1}}
\newcommand{\runze}[1]{\textcolor[rgb]{0.5,0.8,0}{RUNZE: #1}}
\newcommand{\zsd}[1]{\textcolor[rgb]{0.5,0.1,0.8}{ZSD: #1}}
\newcommand{\ashwin}[1]{\textcolor[rgb]{0,0.8,0.4}{ASHWIN: #1}}

\fi

\else
\newcommand{\hide}[1]{}
\newcommand{\note}[1]{}
\newcommand{\added}[1]{#1}
\newcommand{\modif}[1]{#1}
\newcommand{\cut}[1]{}

\newcommand{\todo}[1]{}

\newcommand{\deleted}[1]{}

\newcommand{\temporary}[1]{}

\newcommand{\nuwan}[1]{}
\newcommand{\runze}[1]{}
\newcommand{\zsd}[1]{}
\newcommand{\ashwin}[1]{}

\fi

\newcommand{\prepilotstudy}[1]{\textit{pre-pilot#1}}
\newcommand{\duringpilotstudy}[1]{\textit{during-pilot#1}}
\newcommand{\postpilotstudy}[1]{\textit{post-pilot#1}}

\newcommand{\Prepilotstudy}[1]{\textit{Pre-pilot#1}}
\newcommand{\Duringpilotstudy}[1]{\textit{During-pilot#1}}
\newcommand{\Postpilotstudy}[1]{\textit{Post-pilot#1}}

\newcommand{\DesignGoalOne}[0]{\textit{Support Observations in Situated Contexts}}
\newcommand{\DesignGoalTwo}[0]{\textit{Reduce Task Load of Experimenters}}
\newcommand{\DesignGoalThree}[0]{\textit{Expedite Data Recording, Analysis, and Generation of Creative Insights}}
\newcommand{\Annotation}[1]{\textit{Annotation#1}}
\newcommand{\ScreenshotAnnotation}[0]{\textit{Screenshot}}
\newcommand{\FocusAnnotation}[0]{\textit{Focus}}
\newcommand{\TrueAnnotation}[0]{\textit{Correct}}
\newcommand{\FalseAnnotation}[0]{\textit{Incorrect}}
\newcommand{\CounterAnnotation}[0]{\textit{Counter}}
\newcommand{\VoiceAnnotation}[0]{\textit{Voice}}
\newcommand{\WizardingInterface}[0]{\textit{Wizarding Interface}}

\newcommand{\Menu}[1]{``{#1}''}
\newcommand{\ShortcutKey}[1]{``{#1}'' key}

\newcommand{\analyzer}[0]{\textit{Analyzer}}

\newcommand{\Mary}[0]{Mary}
\newcommand{\direct}[0]{\emph{direct}}
\newcommand{\indirect}[0]{\emph{indirect}}
\usepackage{tcolorbox}

\newcommand{\scenario}[1]{%
\begin{tcolorbox}[opacityframe=0, left=0.1mm, right=0.1mm, top=0mm, bottom=0mm] %
\textit{#1} %
\end{tcolorbox}%
\vspace*{-3.4mm} %
}

\usepackage{subcaption}
\usepackage{multirow}
\usepackage{xcolor}
\usepackage{enumitem}
\usepackage{xspace}
\usepackage{booktabs}

\usepackage[nomessages]{fp}% http://ctan.org/pkg/fp (math)
\usepackage{soul} % text highlighting
\usepackage{pifont}
\usepackage{longtable}
\usepackage{makecell}

\usepackage{arydshln} % add dash lines

\usepackage{hyperref} % cross-reference with custom text

\usepackage{tablefootnote}

\newlength\maxlen
\def\databarlength{xx.xx} % 4 digits
\begin{document}

\title[Demonstrating \WOzTool{}]{Demonstrating \WOzTool{}: A Tool to Assist Wizard-of-Oz Pilot Studies with OHMD}

\author{Nuwan Janaka}
\email{nuwanj@u.nus.edu}
\orcid{0000-0003-2983-6808}

\affiliation{%
  \department{Synteraction Lab}
  \institution{Smart Systems Institute, National University of Singapore}  
%   \city{Singapore}
  \country{Singapore}
}

\author{Runze Cai}
\orcid{0000-0003-0974-3751}
\email{runze.cai@u.nus.edu}

  \affiliation{%
  \department{Synteraction Lab}
  \institution{School of Computing, National University of Singapore} 
  \country{Singapore}
}

\author{Shengdong Zhao}
\authornote{Corresponding Author.}
\email{shengdong.zhao@cityu.edu.hk}
\orcid{0000-0001-7971-3107}

\affiliation{%
\department{Synteraction Lab}
\institution{School of Creative Media \& Department of Computer Science, City University of Hong Kong}
\city{Hong Kong}
  \country{China}
}

\author{David Hsu}
\email{dyhsu@comp.nus.edu.sg}
\orcid{0000-0002-2309-4535}

\affiliation{%
  \institution{School of Computing, National University of Singapore} 
  \institution{Smart Systems Institute, National University of Singapore}
  \country{Singapore}
}

\renewcommand{\shortauthors}{Janaka et al.}

% one sentence motivation
% what was done
% what was found (specific)

\begin{abstract}
While pilot studies help to identify potential interesting research directions, the additional requirements in AR/MR make it challenging to conduct quick and dirty pilot studies efficiently with Optical See-Through Head-Mounted Displays (OST HMDs, OHMDs). To overcome these challenges, including the inability to observe and record in-context user interactions, increased task load, and difficulties with in-context data analysis and discussion, we introduce \WOzTool{} \added{(\url{https://github.com/Synteraction-Lab/PilotAR})}, a tool designed iteratively to enhance AR/MR pilot studies, allowing live first-person and third-person views, multi-modal annotations, flexible wizarding interfaces, and multi-experimenter support. 

\end{abstract}




% A "teaser" image appears between the author and affiliation
% information and the body of the document, and typically spans the
% page.
\begin{teaserfigure}
  \includegraphics[width=\textwidth]{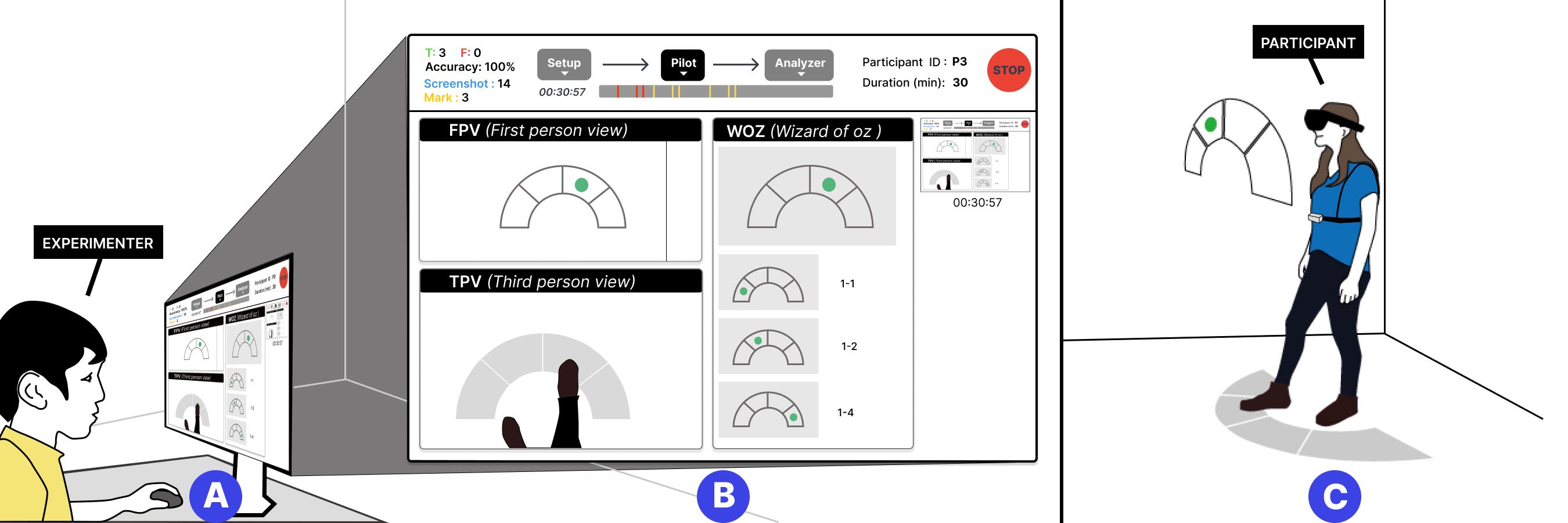}
  \caption{
  (A) The experimenter employs \WOzTool{}, a desktop-based experimenter tool, for OHMD-based pilot studies. 
  (B) \WOzTool{} facilitates real-time monitoring of participants' experiences from both first-person and third-person perspectives, enabling experimenters to track ongoing studies dynamically. In addition, the tool's annotation features allow for the precise marking and capture of significant moments in a photo or video format. Quickly logging quantitative metrics, such as event time, can be done using shortcut keys. Furthermore, a real-time summary of the observed moments and recorded data, available for post-study interviews, promotes in-depth discussions, insights, and support for collaborative review and interpretation. 
  (C) In a separate room, the participant interacts with the simulated AR system, maintaining communication with the experimenter. }
  \Description{The figure consists of two subfigures on the left side, labeled A and B, and one subfigure on the right side, labeled C. Subfigure A shows the experimenter using PilotAR to conduct an OHMD-based pilot study. Subfigure B displays the PilotAR interface, which the experimenter can use to check participants' real-time experiences through First Person View and Third Person View. The experimenter can also control the WizardingInterface to conduct pilots and record important interaction moments using Annotations for post-study review and analysis. Subfigure C shows several IoT devices, such as a fan and a light, and a participant interacting with the simulated system.}
  \label{fig:teaser}
\end{teaserfigure}

\maketitle

% INTRODUCTION (deliver motivation, a bit into second page)
% State of the World
% The big BUT .... (stated as what matters to people)
% Therefore, we did....
% The key findings are...
% The contributions of the work are... (1 to 3)

% RELATED WORK (differentiation, delivered by themes)
% non-defensive, but differentiating
% - Area 1
% - Area 2
% - Area 3

\section{Introduction and Related Work}

Quick and dirty pilot studies validate research concepts, identify usability issues, and guide design decisions without extensive resource commitments \cite{van_teijlingen_importance_2001, thabane_tutorial_2010}. 
However, conducting pilot studies in Augmented Reality (AR) and Mixed Reality (MR) using optical see-through head-mounted displays (OST-HMD, OHMD, or AR smart glasses) poses significant challenges \cite{rey_tool_2021, rey_lopez_ixciimmersive_2022} due to their unique characteristics \cite{janaka_glassmessaging_2023} such as personal, near-eye displays. 

Compared to traditional studies on 2D UIs in desktop/mobile, which mainly observe users from third-person perspective, AR/MR requires both observations from a first-person perspective to understand users' interactions with digital content and a third-person perspective to understand user interactions with the physical world \cite{rey_tool_2021, bellucci_welicit_2021, speicher_what_2019}. Besides observing a multifaceted environment, the task load for experimenters involved in AR/MR pilot studies can also be increased by the need to optionally perform wizard-of-Oz tasks \cite{dow_wizard_2005b, dow_wizard_2005a, de_sa_mobile_2012, bellucci_welicit_2021}, thus necessitating methods to reduce their multitasking burden \cite{bellucci_welicit_2021, brookes_studying_2020, freitas_systematic_2020}. Furthermore, there is insufficient support for in-context data analysis \cite{benford_can_2006, nebeling_mrat_2020, dey_systematic_2018} during the pilot studies, especially for quantitative data, which are typically collected in an informal and raw way. This hinders real-time analysis and deeper discussions in post-study interviews.

Given the absence of an integrated solution for AR/MR pilot studies, despite the development of many specialized tools for individual steps in experiments (e.g., content authoring  \cite{nebeling_trouble_2018, gandy_designers_2014}, rapid prototyping \cite{freitas_systematic_2020}, gesture interaction \cite{ye_progesar_2022, wang_gesturar_2021, lee_wizard_2008}, experiment setup \cite{nebeling_mrat_2020, dey_systematic_2018}, video analysis \cite{zimmerman_observer_2009, kipp_anvil_2014}, 3D and MR visualization \cite{hubenschmid_relive_2022, buschel_miria_2021}, immersive experiment environments \cite{rey_tool_2021, rey_lopez_ixciimmersive_2022, bellucci_welicit_2021}), we created \WOzTool{} (See Appendix~\ref{appendix:comparison}-Table~\ref{tab:related_work:toolkit_comparison} for comparison). It offers experimenters the flexibility to use familiar prototyping or wizarding interfaces rather than requiring the construction of an immersive system with specific skill sets (e.g., \cite{nebeling_mrat_2020, rey_tool_2021, bellucci_welicit_2021} requires Unity3D background), during the early stage of research. Similar to Momento \cite{carter_momento_2007}, \WOzTool{} supports the entire study conduction life cycle: setting up, experimentation, analysis and summarizing, and repeating. However, \WOzTool{} caters to unique challenges of OHMDs (e.g., context, interface \cite{xu_xair_2023, janaka_glassmessaging_2023}), including multiple observation viewpoints real-time synchronization, which is not supported by Momento \cite{carter_momento_2007} as it focuses on applications on mobile phones and desktops.

\WOzTool{} (Fig~\ref{fig:teaser}) is an open-source desktop-based tool for experimenters to conduct AR/MR iterative pilot studies with OHMDs. It streamlines the pilot process from situated observations to results sharing. It incorporates first-person and third-person video observations to help experimenters understand users' in-situ relationship with visual content and environment in real-time and automatically record them for post-analysis. 
It enables annotations, allowing manual or automatic tagging of significant events during the experiment to prevent tedious post-study analysis and missing labeling. \WOzTool{} also allows for task distribution among multiple experimenters, reducing multitasking load and making remote monitoring possible.
Finally, \WOzTool{} enables real-time data summaries, encouraging a deeper discussion during post-pilot interviews and facilitating results sharing with collaborators by exporting data report.
\added{For detailed evaluation, please refer to our original paper, \WOzTool{}~\cite{janaka_pilotar_2024}.}

\section{\WOzTool{} Tool}
\label{sec:tool}

\begin{figure}[hptb]
    \centering
    \includegraphics[width=1\linewidth]{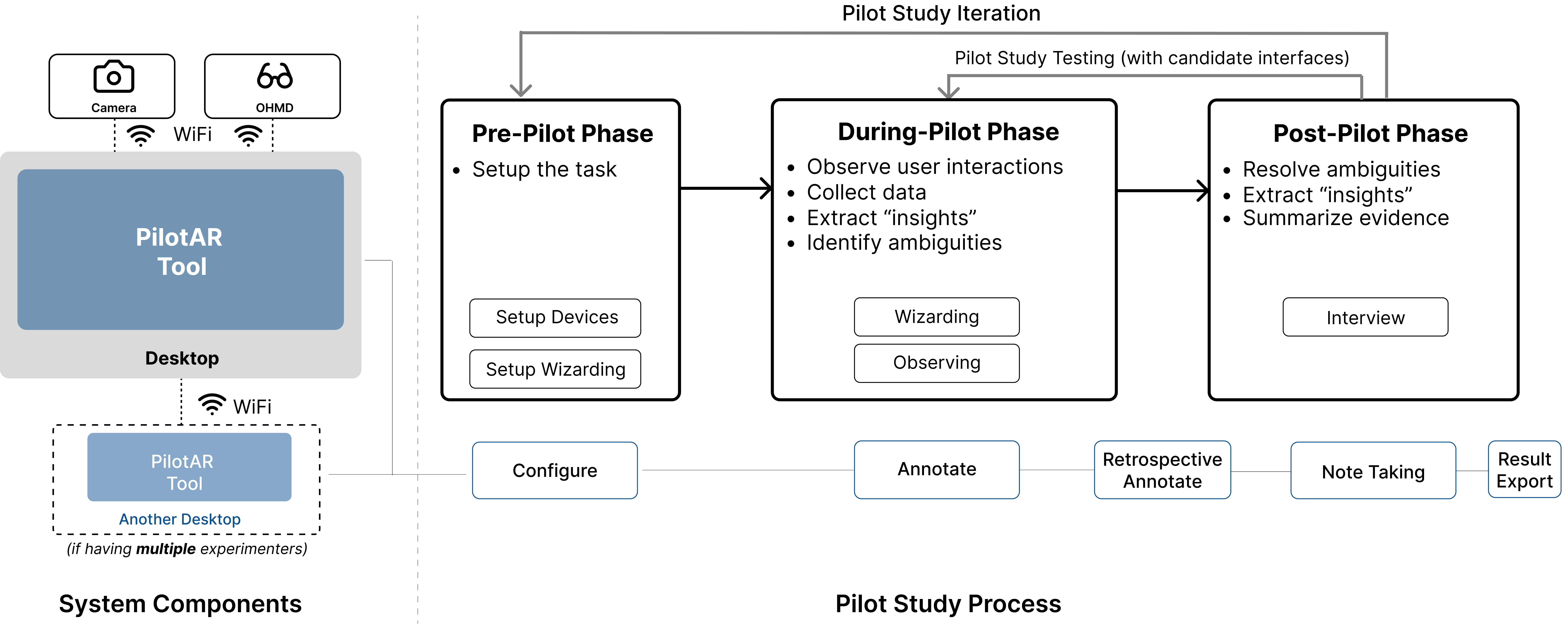}
    \caption{Overview of the system components and workflow with \WOzTool{}.}
    \Description{The diagram presents an overview of the system components and workflow with the PilotAR Tool. The system components involve a desktop connected to a camera and an OHMD via WiFi. There is an option to connect another desktop if multiple experimenters are involved. The workflow diagram outlines steps like configuration, annotation, retrospective annotation, note-taking, and result export. It includes three main phases: Pre-Pilot Phase, During-Pilot Phase, and Post-Pilot Phase. In the Pre-Pilot Phase, tasks such as setting up devices and wizarding are performed. During the Pilot Phase, user interactions are observed, data is collected, insights are extracted, and ambiguities are identified. In the Post-Pilot Phase, ambiguities are resolved, insights are further extracted, and evidence is summarized.}
    \label{fig:tool:overview}
\end{figure}

In this section, we outline the functions and a typical usage scenario of \WOzTool{} (Figure~\ref{fig:tool:overview}). \WOzTool{} integrates features that streamline processes and support replication and innovation in AR/MR pilot studies using the wizard-of-oz \cite{freitas_systematic_2020, alce_wozard_2015, lee_wizard_2008, dow_wizard_2005b, bellucci_welicit_2021} approach. See Appendix~\ref{appendix:implementation} for implementation details.

\subsection{Major Functions}
\label{sec:tool:major_functions}

\textbf{\textit{FPV and TPV Live Streaming (\DesignGoalOne{})}}:
Although relatively straightforward in design, we enabled experimenters to observe participants wearing OHMD in situated contexts through the live first-person view with grids (FPV) and third-person view (TPV), as depicted in Figure~\ref{fig:teaser}. Simultaneous video recorded for subsequent analysis is enabled. Specifically, FPV streams the overlay of digital content and the realistic environment rendered by the OHMD. TPV streams video from a user-attached camera or one positioned by experimenters.
% Both FPV and TPV can be activated from a workflow panel, and their configuration, such as IP address, can also be customized in the UI.

\textbf{\textit{\Annotation{s} with Function Shortcuts (\DesignGoalTwo{})}}:
To facilitate important information documentation during pilot study observations, we enable a variety of annotations. These encompass \ScreenshotAnnotation{} (to capture the screen, optionally with a colored block highlighting a specific Region of Interest (ROI)), \FocusAnnotation{} capturing only a selected screen region), \TrueAnnotation{} and \FalseAnnotation{} (for accuracy calculations), and \CounterAnnotation{} (for tracking interaction attempts). The communication between experimenters and participants is recorded and transcribed to \VoiceAnnotation{} \Annotation{} in text format. 
During pilot studies, experimenters can use customized keyboard shortcuts to activate \Annotation{} functions. These shortcuts can be mapped to UI, user, or experimenter actions for automatic annotations. Additionally, each \Annotation{}'s color can be customized for easy identification, and all annotations are time-stamped for later review.

\textbf{\textit{Multi-experimenter Support (\DesignGoalTwo{})}}:
To reduce task load during pilot studies, we support multi-experimenter scenarios alongside traditional single-experimenter setups. In a single-experimenter scenario, the experimenter concurrently manipulates the wizarding interface, conducts observations, and makes annotations. In the multi-experimenter configuration, one experimenter can act as the wizard, adjusting the interface based on users' actions observed via FPV and TPV, and another experimenter can focus solely on observation and annotation. After the pilot, annotations from both experimenters can be synchronized.

\textbf{\textit{Analyzer (\DesignGoalThree{})}}:
To allow experimenters to get a real-time summary of the collected data, we implemented the \analyzer{} view. By reviewing the annotation index on the recording's timeline, experimenters can identify key moments and use video playback to assist participants in recalling their experiences. Experimenters can adjust annotations recorded during the pilot session (e.g., change timestamp, modify manipulation correctness, modify notes), add new notes, and take screenshots. The analyzer also briefly summarizes accuracy and the time duration between two indices of \Annotation{} and corresponding events.

\textbf{\textit{Summary Review (\DesignGoalThree{})}}:
% \subsubsection{Summary Review (\DesignGoalFourShort{})}
A comprehensive review of the pilot results can be exported from the analyzer to facilitate information sharing among collaborators, including overall descriptive statistics, selected annotation timestamps, notes, and screenshot images. Raw data (e.g., video) can be shared for subsequent analyses.

\subsection{\WOzTool{} Usage Scenario}
\label{sec:tool_usage}

Experimenters might adopt various strategies with \WOzTool{}. Here, we outline a basic approach for conducting a pilot study using \WOzTool{}, with the replication of `Mind the Tap' \cite{muller_mind_2019} as an example to highlight its usage. 

\scenario{\Mary{}, an AR researcher, conceives a novel idea employing foot-tapping as an input interaction for OHMDs  (\textbf{Figure~\ref{fig:teaser}}). She identifies two potential interactions: \direct{} (i.e., the menu appears on the floor within leg's reach) and \indirect{} (i.e., the menu displays in front of the eyes, requiring users to use proprioception to associate it with their foot, Figure~\ref{fig:teaser}C). She aims to discern the strengths and limitations of each foot-tap interaction. Choosing a within-subject design for an initial comparison, \Mary{} opts to employ the wizard-of-oz technique to minimize developmental efforts in a tangible system (e.g., Unity development with optical tracking) and to persuade colleagues to explore this concept further.}

\subsection{Interface and Workflow}
\label{sec:tool:interface_workflow}

The main workflow using \WOzTool{} is divided into three phases: \prepilotstudy{}, \duringpilotstudy{}, and \postpilotstudy{}. This section demonstrates how \Mary{} can utilize \WOzTool{}'s interfaces throughout these phases.

\subsubsection{\Prepilotstudy{} Phase}
See Appendix~\ref{appendix:setup} for details of setup UI.

\scenario{\Mary{} quickly crafts a wizarding interface using Google Slides with a 2x4 menu, where the target location randomizes on subsequent slides. She mirrors these slides to the HoloLens 2 (HL2) via Google Meet on a browser. She uses a phone camera as the TPV by linking it to Google Meet. For \direct{} interactions, the mirrored WOz interface is fixed on the floor. Conversely, for \indirect{} interactions, it's positioned in front of the users' eyes.}

\scenario{\Mary{} initiates the \WOzTool{}, selects `Single User' (Figure~\ref{fig:tool:setup_ui}A), and sets up the devices (Figure~\ref{fig:tool:setup_ui}B) with the HL2 IP address for FPV, a Google Meet link for TPV, and Google Slides for the \WizardingInterface{} (Figure~\ref{fig:tool:setup_ui}C1). She then adds a ``Check foot visibility'' checklist item (Figure~\ref{fig:tool:setup_ui}C2) to verify the FPV setup is accurate before each pilot session. To ascertain accuracy and usability, she enables (Figure~\ref{fig:tool:setup_ui}C3) \TrueAnnotation{}, \FalseAnnotation{}, \CounterAnnotation{}, and \ScreenshotAnnotation{} annotations.}

\subsubsection{\Duringpilotstudy{} Phase}
\label{sec:tool:piloting_ui}

\begin{figure}[hptb]
    \centering
    \includegraphics[width=1\linewidth]{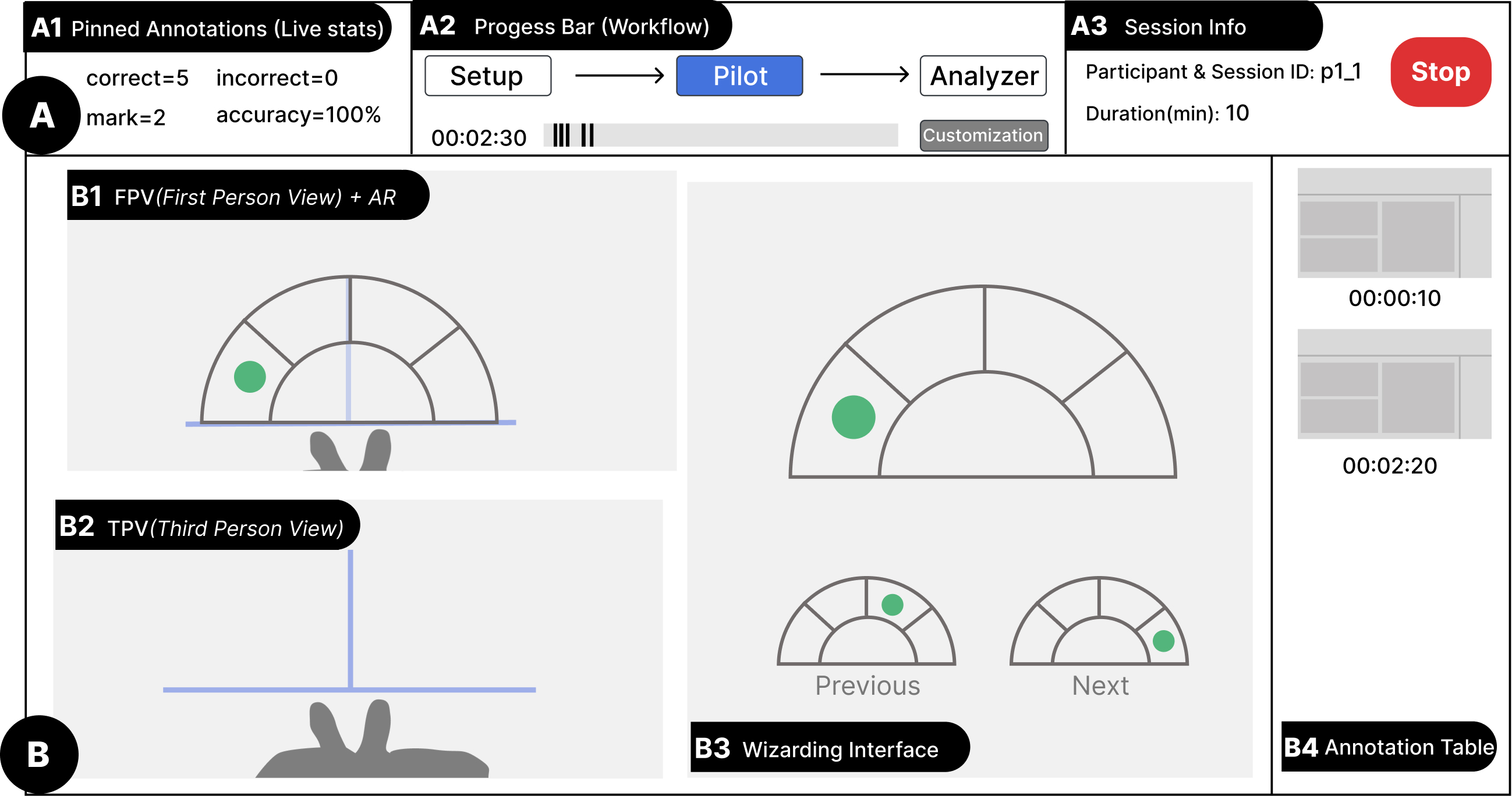}
    \caption{Pilot interface, which includes two major areas. Area (A) is the Top Bar showing (pinned) \Annotation{s}' live statistics (A1), the session progress (A2), and session information (A3). Area (B) presents the main working panel housing the FPV (B1, which shows the digital interface and user's feet from their FPV), TPV (B2), \WizardingInterface{} (B3), and a sidebar for the annotation table (B4).}
    \Description{The figure shows the Pilot interface divided into two major areas. Area A, the Top Bar, includes:
    A1: Live statistics of pinned annotations such as correct, incorrect, marks, and accuracy.
    A2: A progress bar indicating the workflow status through Setup, Pilot, and Analyzer phases, with a time indicator.
    A3: Session information including participant ID, session ID, and a Stop button.
    Area B is the main working panel which includes:
    B1: FPV (First Person View) showing the user's view and digital interface with annotations.
    B2: TPV (Third Person View) providing an external view of the user's position.
    B3: Wizarding Interface showing control options and annotation navigation (Previous, Next).
    B4: A sidebar for the annotation table displaying different time-stamped annotations.}
    \label{fig:tool:piloting_ui}
\end{figure}

After setting up and confirming the checklist, experimenters can enter the anticipated duration and participant and session ID, and initiate the \Menu{Pilot} phase by clicking the \Menu{Start} button (Figure~\ref{fig:tool:piloting_ui}A4).

\textit{Top Bar (Figure~\ref{fig:tool:piloting_ui}A)}.
The top bar displays session-related metadata, including live statistics of measures (e.g., count of Annotations, Figure~\ref{fig:tool:piloting_ui}A1), session progress (e.g., duration and timeline,  Figure~\ref{fig:tool:piloting_ui}A2), and session information (e.g., participant info, anticipated duration,  Figure~\ref{fig:tool:piloting_ui}A3). Experimenters will receive a notification when the anticipated time has elapsed and can stop the session by clicking the \Menu{Stop} button located at the right corner of the top bar.

\textit{Main Working Panel (Figure~\ref{fig:tool:piloting_ui}B)}.
The working panel displays FPV (Figure~\ref{fig:tool:piloting_ui}B1), TPV (Figure~\ref{fig:tool:piloting_ui}B2), and \WizardingInterface{s} (Figure~\ref{fig:tool:piloting_ui}B3), with a layout that can be customized according to the experimenter's preferences. 
In the right corner of the working panel, the captured \ScreenshotAnnotation{} and \FocusAnnotation{} annotations using keyboard shortcut keys (e.g., \ShortcutKey{3} key) are shown as images with timestamps in the Annotation Table (see Figure~\ref{fig:tool:piloting_ui}B4). Clicking on these images opens a pop-up window, allowing the experimenter to add notes to the annotations.

\scenario{
[{Piloting with the First Interface}] \Mary{} starts the pilot with \direct{} interface (Figure~\ref{fig:teaser}C). Adjusting the target location on the \WizardingInterface{} (Figure~\ref{fig:tool:piloting_ui}B3), she annotates accuracy across ten trials, taking screenshots of any interesting behavior (Figure~\ref{fig:tool:piloting_ui}B4). \Mary{} also monitors the trial count and accuracy via the live statistics dashboard (Figure~\ref{fig:tool:piloting_ui}A1).}

\subsubsection{\Postpilotstudy{} Phase}
\label{sec:tool:analyzer_ui}
\begin{figure}[hptb]
    \centering
    \includegraphics[width=1\linewidth]{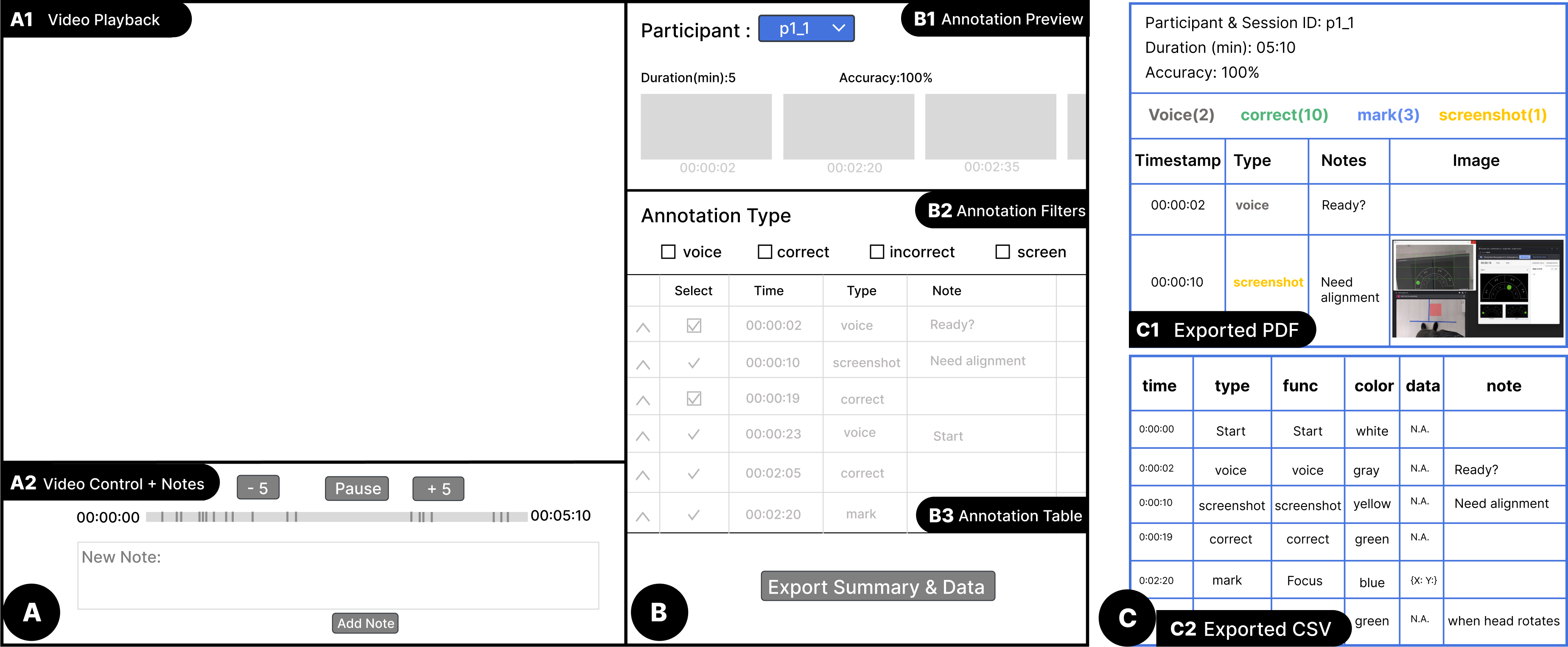}
    \caption{The Analyzer interface comprises two main panels: the video panel (A) and the annotation panel (B). The video panel includes video playbacks of the pilot (A1), video controls, and a new note panel (A2). The annotation panel features an annotation preview (B1), annotation filtering options (B2), an annotation table (B3), and an exporting button. The Analyzer supports exporting the annotations (C) in PDF format (C1) and CSV format (C2).}
    \Description{The figure depicts the Analyzer interface and Exported Summary components. The UI of the Analyzer is shown on the left and center, with:
    Area A on the left, which includes:
    A1: Video playback of the pilot session.
    A2: Video control and new note panel, featuring playback controls and a section to add new notes.
    Area B in the center, which includes:
    B1: Annotation preview showing a summary of annotations.
    B2: Annotation filtering options allowing the user to filter annotations by type (voice, correct, incorrect, screen).
    B3: Annotation table listing annotations with details such as time, type, and notes.
    On the right side of the figure, Area C shows the Exported Summary in:
    C1: PDF format, including participant information, session details, and an overview and specifics of annotations.
    C2: CSV format, detailing time-stamped annotations with attributes like type, function, color, data, and notes.}
    \label{fig:tool:analyzer_ui}
\end{figure}

Upon completion of the pilot session, the \analyzer{} window appears (Figure~\ref{fig:tool:analyzer_ui}), displaying the video panel on the left (Figure~\ref{fig:tool:analyzer_ui}A) and \Annotation{s} panel on the right (Figure~\ref{fig:tool:analyzer_ui}B).

\textit{Video Panel(Figure~\ref{fig:tool:analyzer_ui}A)}. 
It can play the recorded video (Figure~\ref{fig:tool:analyzer_ui}A1) and navigate to any timestamp by clicking the timeline (Figure~\ref{fig:tool:analyzer_ui}A2) or using three buttons to rewind, pause, and fast-forward. Experimenters can create new \Annotation{s} with notes in the \Menu{New Note} area below the video timeline (Figure~\ref{fig:tool:analyzer_ui}A2).

\textit{\Annotation{} Panel (Figure~\ref{fig:tool:analyzer_ui}B)}.
It features an annotation preview (Figure~\ref{fig:tool:analyzer_ui}B1), annotation filtering options (Figure~\ref{fig:tool:analyzer_ui}B2), an annotation table (Figure~\ref{fig:tool:analyzer_ui}B3), and an exporting button. The annotation preview (Figure~\ref{fig:tool:analyzer_ui}B1) provides an overview of the pilot, including its duration, manipulation accuracy, and collected screenshots. Experimenters can click on these screenshots to pinpoint annotated moments in the recorded video.

Within the Annotation table (Figure~\ref{fig:tool:analyzer_ui}B3), experimenters have the capability to view and adjust annotation details by double-clicking on a cell. Additionally, specific Annotations can be highlighted by clicking the corresponding icon in the first column or applying the filters available (Figure~\ref{fig:tool:analyzer_ui}B2). The tool also facilitates the export of summaries and selected \Annotation{s} in both PDF and CSV formats (Figure~\ref{fig:tool:analyzer_ui}C).

\scenario{[Analysis]
Upon finishing the session, the \analyzer{} activates, presenting screenshots, accuracy data, and annotations (Figure~\ref{fig:tool:analyzer_ui}). Before the interview, Mary reviews these annotations and accuracy (Figure~\ref{fig:tool:analyzer_ui}B1-B3), devising questions for further inquiry. For clarity on specific screenshots, she replays footage from 5 seconds prior (Figure~\ref{fig:tool:analyzer_ui}A1-A2). She then conducts the interview, discussing the participant's experiences and challenges, and incorporates their feedback into the annotation notes (Figure~\ref{fig:tool:analyzer_ui}B3).}

Experimenters can return to the \Menu{Pilot} session for subsequent pilot studies and initiate new recordings. All interactions in the \analyzer{} are stored, enabling experimenters to switch between different pilot recordings using the drop-down menu in Figure~\ref{fig:tool:analyzer_ui}B1.

\scenario{[Piloting with the Alternative Interface] 
After assessing the \direct{} interface, \Mary{} tests the \indirect{} interface in the same approach.} 

\scenario{[Overall Analysis] 
After piloting both interfaces, \Mary{} invites the participant for an overall interview, utilizing the \analyzer{} to toggle between pilot recording sessions or view them simultaneously (Figure~\ref{fig:tool:analyzer_ui}B1). This comparison offers insights into ``rough'' accuracy and usability variations, which are noted in \analyzer{} (e.g., \textit{direct} one is slightly more accurate while causing neck pain for long usage, (Figure~\ref{fig:tool:analyzer_ui}B3).}

\scenario{[Repeating] \Mary{} replicates this process with three more participants, counterbalancing the interface. \Mary{} exports participant data summaries in PDF (Figure~\ref{fig:tool:analyzer_ui}C1) and shares them with colleagues to convince the differences between \direct{} and \indirect{} interfaces. She cites participant feedback and replays specific recordings for context when queried for details.}

\scenario{[Further Exploration: Multi-experimenter]
Seeing the team's interest, \Mary{} broadens their exploration to assess how interaction accuracy and speed vary between two interfaces as menu size changes. She trains a colleague to act as the wizard, thus reducing the wizarding workload and focusing more on observations. After creating additional slides for varied menu sizes (e.g., 1x2, 2x4, 3x6), they conduct pilot tests with four participants using a between-subjects design. To calculate the speed of interactions, they combine \TrueAnnotation{}/\FalseAnnotation{} annotations with custom annotations that automatically mark target changes (linked to slides' changes).
After each pilot session, data is exported to CSV (Figure~\ref{fig:tool:analyzer_ui}C2) for graph generation in Excel, which facilitates comparing relationships among speed, accuracy, and menu size. Convinced that their pilot study has uncovered a notable trend, the team decides to transition to a formal study.}

\scenario{[Summary] Employing the wizard-of-oz methodology with \WOzTool{}, the team expedites (e.g., less than one week as opposed to a full-fledged motion tracking application, which can take several weeks to months) the identification of viable research directions. Using \WOzTool{}, experimenters can overcome challenges in rapidly evaluating diverse concepts, gathering preliminary quantitative measures for comparison, and convincing colleagues, significantly shortening the knowledge discovery phase.}

% DISCUSSION (what was interesting, what matters)

% What are the implications of your results?
% What do they mean for this topic and your field?
% What is important and worthy of again being called to the reader's attention?
% What was surprising, unexpected, intriguing?
% Did you fulfill the promises you set out in the Intro via the claims of your work?
% Were any hypothesis confirmed or disconfirmed?
% Were the predictions of any theory upheld or refuted?
% What worked and what did not work? And why?

% - add limitations here to guide interpretation of work

% CONCLUSION
% affirm that you have delivered on the claims made in your Introduction
% Summarize the contributions of the work
% Make any key points with which you would like to leave the reader, 
% point to a bright future/better world for your work having been done in it. 

% FUTURE WORK (what are the “big idea” next steps to follow from your work?)
% Avoid merely incremental steps like a todo list

% Try to frame the contributions of the work such that they speak to your broader scholarly community, not just those interested in your narrow topic:

\section{Conclusion}

As AR/MR technology is poised to shape the future immersive world, including the metaverse, facilitating interactions between digital and physical entities becomes paramount. This underscores the importance of tools tailored for refining these interactions through pilot studies. As an initial step, we introduce \WOzTool{}, an open-source tool (\modif{\url{https://github.com/Synteraction-Lab/PilotAR}}) designed to support such studies. It enables real-time and retrospective multi-viewpoint observations, notes, and filters of crucial observations, thus facilitating comprehensive discussions with participants and researchers to discover insights effectively. Additionally, it can share the pilot study process, data, and insights with the larger research community. Its all-in-one capability can be applied as a standalone observation tool or a video analyzer tool to border studies beyond pilot studies or OHMD-based studies.

\begin{acks}
We would also like to thank Tan Si Yan and Siddanth Ratan Umralkar for developing specific system components. Additionally, we wish to thank the anonymous reviewers for their valuable time and insightful comments.
This research is supported by the National Research Foundation, Singapore, under its AI Singapore Programme (AISG Award No: AISG2-RP-2020-016). The CityU Start-up Grant 9610677 also provides partial support.
Any opinions, findings, conclusions, or recommendations expressed in this material are those of the author(s) and do not reflect the views of the National Research Foundation, Singapore. 

\end{acks}

\bibliographystyle{ACM-Reference-Format}
\bibliography{paper/references}

\appendix

\section{Comparison}
\label{appendix:comparison}

Table~\ref{tab:related_work:toolkit_comparison} highlights the differences and similarities between \WOzTool{} and prior tools.

\begin{table*}[htbp]
\centering
\caption{Summary of the feature comparisons between the tools for conducting AR-related studies. Here, FPV = first-person view, TPV = third-person view, AR = virtual content view. Note: This list is not
exhaustive. Although DART \cite{gandy_designers_2014, macintyre_dart_2004} is meant for authoring AR/MR content, we have added it here for comparison as it supports various functions that could also be used for conducting AR/MR experiments.}
\label{tab:related_work:toolkit_comparison}
\small
\begin{tabular}{@{}p{0.15\textwidth}p{0.15\textwidth}p{0.15\textwidth}p{0.15\textwidth}p{0.13\textwidth}p{0.15\textwidth}@{}}
\toprule
\textbf{Tool/Toolkit} & Lee et al. \cite{lee_wizard_2008} & 
Rey et al. \cite{rey_tool_2021, rey_lopez_ixciimmersive_2022} (IXCI) & 
MacIntyre et al. \cite{gandy_designers_2014, macintyre_dart_2004} (DART) &
Nebeling et al. \cite{nebeling_mrat_2020} (MRAT) &
Proposed tool (\textbf{\WOzTool{}}) \\ \midrule
\textbf{Purpose} & 
Identify multimodal inputs for AR manipulation tasks and how AR display conditions affect them &
Support research by streamlining immersive user studies &  
An authoring tool enabling rapid prototyping of AR applications by designers/non-technologists &  
An experimenter support tool for analyzing MR experiences &
An experimenter support tool for conducting AR/MR \textbf{pilots}, data collection, and analysis
\\
\midrule
\textbf{Target studies} &
WOz studies &  
Unity3D-based studies &
AR studies &  
Unity3D-based studies &
Pilot studies in AR/MR, including WOz\\
\midrule
\textbf{Prototype fidelity} &
High &  
High &
Low-High &  
High &
Low-High \\
\midrule
\textbf{Multiple experiment support} &
Single &  
Multiple &  
Multiple &  
Multiple &
Multiple \\
\midrule
\textbf{Observation support} &
FPV, TPV &  
AR &  
FPV, AR, TPV &  
Interaction data-points &
FPV with AR, TPV \\
\midrule
\textbf{Recording support} &
\ding{51} &  
\ding{55} &  
\ding{51} &  
Processed spatial-temporal interaction data points &
\ding{51} \\
\midrule
\textbf{Note taking} &
\ding{55} &  
\ding{55} &  
\ding{55} &  
\ding{55} &
\ding{51} \\
\midrule
\textbf{Post-analysis} &
\ding{55} &  
\ding{55} &  
Not applicable &  
\ding{51} &
\ding{51} \\ 
\midrule
\textbf{Summarizing and exporting} &
\ding{55} &  
\ding{55} &  
Not applicable &  
\ding{51} &
\ding{51} \\
\bottomrule
\Description{This table summarizes comparisons between the tools that assist experimenters in conducting AR/MR-related studies.}
\end{tabular}
\end{table*}

\section{Implementation}
\label{appendix:implementation}

We used Python (3.9) as our primary programming language due to its cross-platform compatibility (e.g., Windows, MacOS). To achieve the tool's functionalities, we incorporated several third-party packages. The user interface (UI) was developed using Tkinter\footnote{\url{https://docs.python.org/3/library/tkinter.html}} and related theme packages, such as CustomTkinter\footnote{\url{https://github.com/TomSchimansky/CustomTkinter}}. 
The \WOzTool{} utilizes Pynput\footnote{\url{https://pypi.org/project/pynput}} to monitor user inputs and FFmpeg\footnote{\url{https://ffmpeg.org}} to handle screen recording. For video playback, we used Python-VLC\footnote{\url{https://pypi.org/project/python-vlc/}} and audio transcription we used Whisper\footnote{\url{https://openai.com/blog/whisper/}}. FFmpeg and websocket were incorporated to enable video and data streaming between the wizard and the observer in multi-experimenter settings. Detailed information about the \textbf{open-source} implementation can be found in \modif{\url{https://github.com/Synteraction-Lab/PilotAR}}.

\section{\WOzTool{} Setup}
\label{appendix:setup}

\begin{figure}[hptb]
    \centering
    \includegraphics[width=1\linewidth]{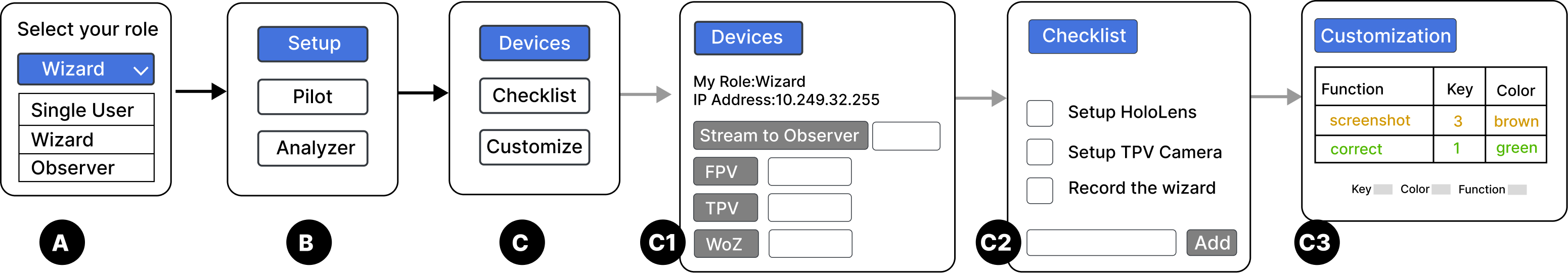}
    \caption{Workflow of Setup UI. Upon starting the tool, the experimenter is prompted to select the role (A), including single- and multi-experimenter (wizard/observer). Then, menu (B) indicates the three major steps of conducting a pilot study: Setup, Pilot, and Analyzer. In Setup (C), there are three sub-steps, including device configurations (C1), checklist configuration (C2), and annotation customization (C3).}
    \Description{The image illustrates the Workflow of the Setup UI for the PilotAR Tool. The workflow starts with the experimenter selecting a role (A) from options like Single User, Wizard, and Observer. Following role selection, the menu (B) displays the main steps for conducting a pilot study: Setup, Pilot, and Analyzer. During the Setup phase (C), there are three sub-steps: 
    1. Device configuration (C1) where the role, IP address, and streaming options (FPV, TPV, WoZ) are set up.
    2. Checklist configuration (C2) which includes tasks like setting up HoloLens, setting up the TPV camera, and recording the wizard.
    3. Annotation customization (C3) where functions are assigned to specific keys with corresponding colors (e.g., "screenshot" is assigned to key 3 with the color brown, and "correct" is assigned to key 1 with the color green).}
    \label{fig:tool:setup_ui}
\end{figure}

\paragraph{Role Selection (Figure~\ref{fig:tool:setup_ui}A)} 
Upon launching the tool, the experimenter is prompted to select their role:  \textit{single-user} for single-experimenter pilots or \textit{wizard/observer} for multi-experimenter pilots.

\paragraph{Device Configuration (Figure~\ref{fig:tool:setup_ui}C1)}
This task allows the experimenter to input essential information such as FPV and TPV connections (e.g., IP address, credentials), \WizardingInterface{} (e.g., Google Slides URL link or python file path), and screen recording inputs (e.g., video and audio source), making them all displayed on the monitor.

\paragraph{Checklist Creation (Figure~\ref{fig:tool:setup_ui}C2)} 
The checklist aids in remembering crucial steps during the pilot study, such as confirming OHMD, TPV camera, and recording. Customizable items can be added by typing in the provided space at the bottom.

\paragraph{Shortcut Key Customization (Figure~\ref{fig:tool:setup_ui}C3)} 
Experimenters can manage which \Annotation{s} are displayed during the pilot session (known as Pinned \Annotation{}) and customize aspects like color, name, and shortcut key.

\end{document}